\documentclass{mem}
\usepackage{natbib}\usepackage{txfonts}\usepackage{balance}
\usepackage{graphicx}
\usepackage[a4paper]{hyperref}
\idline{75}{282}
\begin{document}
\def\teff{$T\rm_{eff }$}
\def\kms{$\mathrm {km s}^{-1}$}

\def\ltsim{\raise 2pt \hbox {$<$} \kern-1.1em \lower 4pt \hbox {$\sim$}}
\def\gtsim{\raise 2pt \hbox {$>$} \kern-1.1em \lower 4pt \hbox {$\sim$}}

\title{
Observational properties of diffuse radio sources in Galaxy Clusters
}

\subtitle{Current knowledge and open questions}

\author{
T. \, Venturi\inst{1}
}

  \offprints{T. Venturi}

\institute{$^1$
Istituto Nazionale di Astrofisica --
Istituto di Radioastronomia, Via Gobetti 101,
I-40131 Bologna, Italy
\\
\email{tventuri@ira.inaf.it}
}

\authorrunning{Venturi}

\titlerunning{Diffuse radio sources in galaxy clusters}

\abstract{

Diffuse radio sources in galaxy clusters, i.e. radio halos and relics, are 
unique signposts of cluster assembly in the Universe. Our knowledge of their 
observational properties has considerably improved over the past decade,
and the long standing questions concerning their origin and rarity are now 
starting to receive some answers. It is nowadays fairly well established 
that massive cluster mergers are the key ingredient to account for the 
origin of halos and relics: they induce shocks and turbulence in the 
cluster volume, whose energy is able to (re)--accelerate relativistic 
electrons in the intracluster medium. 
\\
The increasing number of halos and relics found, coupled with the  high 
quality information available in the X--ray band for a large number of 
clusters, now allow to derive significant statistical correlations between
the radio and X--ray properties of the diffuse sources and the hosting clusters.
It has been shown that galaxy clusters may be {\it radio--loud} or 
{\it radio--quiet} with respect to the presence of radio halos. Morevoer, 
radio halos and relics are present only in unrelaxed clusters, whereas no 
such sources have ever been found in relaxed systems.
It has also become clear that a large distribution exists for the spectral
index of the synchrotron spectrum of radio halos, and sources
with $\alpha$ up to $\sim 2$ (S$\propto\nu^{-\alpha}$) have been recently
found.
\\
Despite all this, a number of observational issues remain open, moreover
our improved knowledge has opened new questions. Well defined radio
spectra, from $\sim$ hundred MHz to GHz frequencies and good spectral
imaging are available only for few halos and relics, and such information 
is crucial for a detailed understanding of their origin.
It is presently unclear if there are basic differences between clusters 
with ultra steep spectrum radio halos and those hosting ``classical'' ones,
and it is unknown if  less energetic mergers result in an observational 
signature in the radio band.
\\
With the forthcoming advent of LOFAR and of the SKA Pathfinders, we will
soon be able to observe galaxy clusters with a major improvement in the
radio sensitivity and frequency coverage, and we expect to be able to 
soon address some of these pressing questions.

}

\maketitle{}

\section{Introduction}
Radio observations are a unique tool to study the formation and evolution 
of galaxy clusters and their constituents. 
The presence of diffuse Mpc scale radio emission in a number of galaxy 
clusters, i.e. {\it radio halos} and {\it relics}, reveals the 
presence of relativistic particles and magnetic fields extending 
throughout the cluster volume, while the bent radio emission associated
with cluster galaxies is the signature of the galaxy motion within 
the clusters and on large scale bulk motion of the intracluster 
medium (ICM).

It is nowadays becoming clear that halos and relics are closely connected to 
the cluster formation history, therefore our understanding of their origin 
and evolution is not only relevant in itself, but it is crucial for our 
global understanding of the mechanisms at play during the processes of 
cluster assembly in the Universe.

In the following I will provide an overview of our current
observational knowledge of the radio emission from galaxy clusters, with 
particular emphasis on radio halos and relics. I will also underline some
of the current urgent open questions, which are expected to receive
an answer once the next generation radio interferometers are fully
operational.

Throughout the paper I will assume S$\propto \nu^{-\alpha}$ and a $\Lambda$CDM
cosmology, with H$_o$=70 km s$^{-1}$Mpc$^{-1}$, $\Omega_{\rm m}=0.3$,
$\Omega_{\rm \Lambda}=0.7$.

\section{Radio emission from galaxy clusters}

\subsection{Radio galaxies}
 
\begin{figure}[htbp!]
\resizebox{\hsize}{!}{\includegraphics[clip=true]{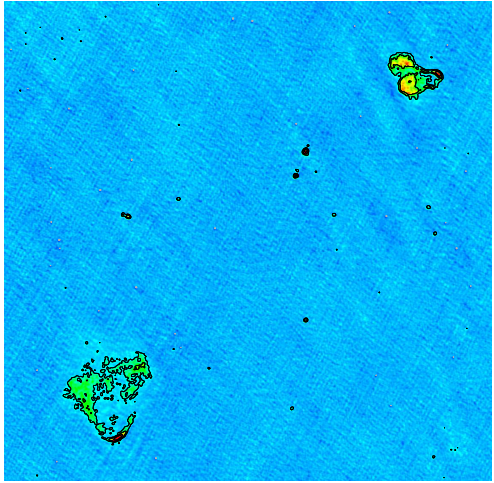}}
\caption{\footnotesize
\textit Wide--angle tail and narrow--angle tail radio galaxies in A\,754.
The observations were carried out with the GMRT at 610 MHz. The resolution of
the image is 5.3$^{\prime\prime}\times4.9^{\prime\prime}$ and the 1$\sigma$ noise
level is 0.1 mJy/b. Contours are $\pm$0.5,2,8 mJy/b.}
\label{fig:fig1}
\end{figure}

The most beautiful examples of radio emission from individual cluster
members are the double--lobed FR\,I 
\citep{fr74}
radio galaxies, whose jets and lobes usually extend well beyond the 
optical light. Their morphology is often misaligned: 
jets and lobes are bent in U or C shape (narrow--angle and wide--angle 
tail radio galaxies respectively), possibly by
the combination of galaxy motion within the cluster and local flows in 
the ICM due to cluster mergers 
\citep{bliton98,burns98,fv02}.
A beautiful example is reported in Fig. \ref{fig:fig1}, which shows a number 
of distorted radio galaxies in A\,754.
Studies of cluster radio galaxies are relevant not only to derive information
on the cluster dynamics and properties of the ICM, but also for the study of 
the cluster magnetic 
field, by means of Faraday rotation and depolarization (see Murgia, present
volume;\citealt{bonafede10};\citealt{govoni10};\citealt{vacca10}).

The radio emission from the central dominant cluster 
galaxy plays a special role. In some cases the location of the radio 
emission from the lobes shows a correlation with the presence
of X--ray features 
(e.g., A\,262\citealt{clarke09};$~$NGC\,5813,\citealt{randall11};$~$HCG\,62 
and other systems,\citealt{giacintucci11a}), which is suggestive of feedback 
between the central radio emission and the
ICM \citep{mnn07}.

\subsection{Mini--halos}

\begin{figure}[htbp!]
\resizebox{\hsize}{!}{\includegraphics[clip=true]{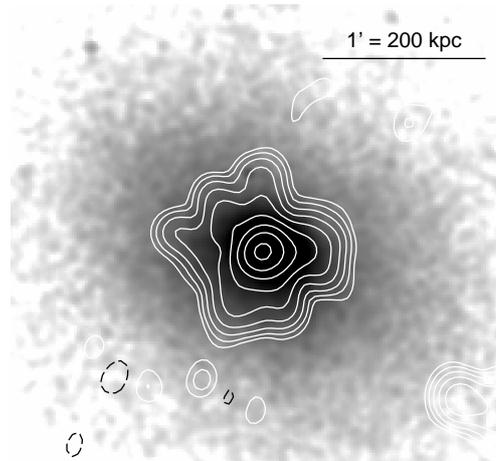}}
\caption{\footnotesize
\textit GMRT 325 MHz contours of the mini--halo in RXCJ\,1504.1--0248
overlaid on the {\it Chandra} X--ray emission. The resolution of the
radio image is 11.3$^{\prime\prime}\times10.4^{\prime\prime}$; contours start at 
$\pm$0.3 mJy/b (3$\sigma$) and are spaced by a factor of two.
See \citealt{giacintucci11b}.}
\label{fig:fig2}
\end{figure}

In very few relaxed clusters, the central radio AGN is surrounded by diffuse 
emission in the form of a halo, with steep spectral index ($\alpha > 1$),
and linear extent of the order of few
hundreds of kpc. A typical mini--halo is reported in Fig. \ref{fig:fig2}.
Only 9 mini--halos are known to date 
(\citealt{giacintucci11b} and references therein), 
and their origin is an issue, since the diffusion 
time of the relativistic electrons is shorter than the travel time necessary
to cover the whole extent of these sources. Cold fronts in the ICM have been
observed in some clusters hosting a mini--halo, suggesting that gas 
sloshing may be responsible for the origin of the cold fronts and may provide 
the turbulence needed to re--accelerate seed electrons, most likely coming 
from the radio activity of the central AGN, to form a mini--halo 
(\citealt{mg08}, ZuHone et al., present volume).
It has been found that radio mini--halos and giant radio halos (see Sect. 2.3 
and 3) share the same correlation between the radio power and X--ray luminosity,
but the two classes of sources are clearly separated in the 
radio power--radio size diagram, indicating that the emissivity of radio
mini--halos is larger than that of giant radio halos
\citep{cassano08a,murgia09}.

%
%
%

\subsection{Halos and relics}
Diffuse radio emission, whose overall size may reach and exceed the Mpc size, 
is observed in a number ($\sim$ 40) of galaxy clusters. Such emission comes 
into two main flavours. {\it Radio halos}, located at the centre of galaxy 
clusters, exhibit a fairly regular morphology, in good spatial coincidence
with the distribution of the hot X--ray emitting gas, and their radio
emission is unpolarized (albeit a couple of noticeable exceptions, see
Pizzo, present volume);
{\it radio relics} may have a variety of morphologies (elongated and 
arc--shaped are the most common), are located at the cluster periphery and 
show high fractional polarization. Their monochromatic radio power at 
1.4 GHz is in the range P$_{\rm 1.4~GHz} \sim 10^{23}-10^{25}$ W Hz$^{-1}$.
Both types of sources have no obvious optical counterpart, suggesting that 
the radio emission is connected with the ICM, and implying the existence of 
relativistic leptons with energy of few GeV (Lorentz factor $\gamma \sim 10^4$)
and $\mu$G magnetic field mixed with the hot (T$\sim$5--10 KeV) ICM
(see\citealt{ferrari08} and\citealt{cassano09} for recent reviews).

From a theoretical point of view, halos and relics
have always been a challenge for our understanding of 
their origin, due to their {\it large extent} and {\it rarity}. The diffusion 
time of relativistic electrons to spread over the Mpc scale size exceeds their 
radiation lifetime by roughly two orders of magnitude ($10^{10}$ years to be 
compared to $\sim 10^8$ years respectively), requiring some form of 
re--acceleration (e.g.,\citealt{jaffe77}).
In the past few years observational evidence has accumulated in favour of the
idea that the main source of electron re--acceleration in galaxy clusters
comes from the injection of shocks and turbulence induced by massive
cluster mergers (e.g.,\citealt{brunetti01},\citealt{bl07},\citealt{vazza09}).
 
\begin{figure}[htbp!]
\resizebox{\hsize}{!}{\includegraphics[clip=true]{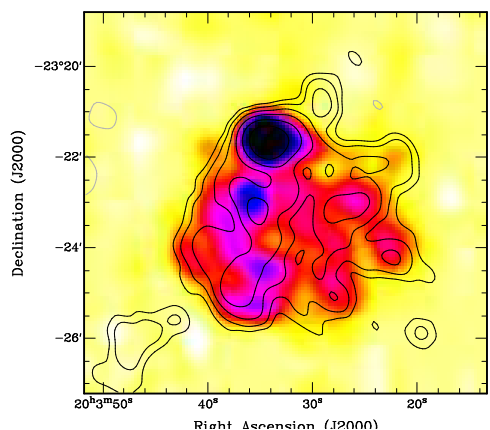}}
\caption{\footnotesize
\textit Radio halo in RXCJ\,2003.5--2323: VLA 1.4 GHz emission (grey scale)
with 240 MHz GMRT contours overplotted. The restoring beam is
35$^{\prime\prime}\times35^{\prime\prime}$ in both images. The first contour
at 240 MHz is $\pm$ 1 mJy/b (2.5$\sigma$) and contours are spaced by a 
factor of 2. The distance of the cluster is z=0.317, and 
1$^{\prime\prime}$=4.62kpc.}
\label{fig:fig3}
\end{figure}

RXCJ\,2003.5--2323 (z=0.317), reported in Fig. \ref{fig:fig3}, is one 
of the many examples of giant radio halos. It is one of the most powerful and 
among the most distant found so far. Its size is $\sim$ 1.4 Mpc, and does 
not change in the frequency range 235 MHz -- 1.4 GHz.
The brightness distribution is patchy at all wavelenghts, but this is 
not always the case. In a number of radio halos the 
surface brightness peaks in the central cluster region (in coincidence
with the X--ray peak) and smoothly decreases at the edges.

The very low surface radio brightness of halos and relics, combined with 
their steep radio synchrotron spectra (average values reported for 
$\alpha$ are in the range 1.2--1.4, see Section 4 below) make 
their detection difficult. 
At present, about 25 radio halos have been imaged at high sensitivity 
(e.g.,\citealt{giovannini09} and references therein, Brunetti et al. 2008,
Bonafede et al. 2009a, Giacintucci et al. 2009a, Clarke \& Ensslin 2005,
Venturi et al. 2003, 2007 \& 2008),
and about 25 relics are known to date (e.g.,\citealt{vw09a} and references 
therein).
A number of clusters host both a radio halo and a relic (i.e. Coma, 
Kim et al. 1989, Brown \& Rudnick 2011; 
A\,521,\citealt{brunetti08};~A\,1300,\citealt{venturi09} 
and Giacintucci et al., present volume;~A\,2255,\citealt{pizzo08}; 
A\,2256,\citealt{ce05};~A\,2744,\citealt{orru07} and references therein, 
Giacintucci et al., present volume). 
A few  clusters with a double relic have 
been recently found (i.e. A\,1240 and A\,2345,\citealt{bonafede09b}; 
A\,548b,\citealt{feretti06}; CIZA J\,2242.8+5301,\citealt{vw10} and
van Weeren et al., present volume; ZwCl\,2341.1+0000,\citealt{vw09b}).
A unique case is the cluster RXCJ\,1314.4--2515, which hosts
two relics and one radio halo \citep{venturi07}.

%

%

This long list of references clearly shows that this field is very hot, and 
that high sensitivity radio imaging is being collected at high rate. 
The amount of high quality information currently available both in the radio 
and in the X--ray band now allows an accurate comparison between the 
observations and the current
models proposed for the origin of halos and relics. In the next Sections 
I will report on some of the  recent most outstanding observational 
achievements in this field.
The papers by Brunetti and Cassano, in this volume, will focus respectively
on the theoretical and statistical side of the origin and evolution
of radio halos.

\section{Radio halo--cluster merger connection and cluster bimodality}

The first hints of a connection between the thermal (X--ray emission from 
the ICM) and non--thermal (radio halos) properties of galaxy clusters
date back about a decade ago, when it was found that 
statistical correlations exist between global cluster properties (such as 
temperature, mass and X--ray luminosity) and the radio power of halos.
In particular, more X--ray luminous (and hence more massive) clusters host 
more powerful radio halos (e.g.,\citealt{giovannini99},\citealt{liang00},\citealt{cassano06}). 
Govoni et al. (2004) 
first found a positive correlation between the local radio brightness and
the X--ray flux density; Buote (2001) first provided a quantitative 
evaluation of the dynamical disturbance of galaxy clusters with  
diffuse radio emission.

The literature information was recently combined with the outcome of 
the 610 MHz GMRT Radio Halo Survey 
(\citealt{venturi07} and{\citealt{venturi08}), to constrain the statistical 
properties of clusters hosting radio halos in the redshift interval z=0--0.4
on much more solid grounds.
\\
It was found that the fraction of clusters hosting a radio halo increases 
with increasing X--ray luminosity, and hence cluster mass, at a 
$\sim4\sigma$ significance level \citep{cassano08b}. 
\\
In order to fully exploit the observational information of the GMRT 
survey, upper limits to the radio power of a radio halo were estimated
for those clusters without detection, and such values were reported on the
same logL$_{\rm X}$--logP$_{\rm 1.4~GHz}$ correlation for radio halos 
\citep{brunetti07}.
Remarkably, the non--detections are not the result of 
limited sensitivity, and the clusters in the sample show a clear bimodal 
behaviour: galaxy clusters either host a radio halo, whose radio power 
correlates with the cluster X--ray luminosity, or have upper limits, which
populate the bottom--right part of the logL$_{\rm X}$--logP$_{\rm 1.4~GHz}$
diagram and are placed ~1.5--2 orders of magnitude
below the correlation. This result supports a ``transient'' nature of radio 
halos and is in agreement with the expectation of the re--acceleration model 
\citep{brunetti07,brunetti09}.
\\
Once accurately scrutinized, clusters with and without halo show clear
differences in terms of X--ray properties. Using the literature information
on the cluster dynamical status,
Venturi et al. (2008) showed that all radio halos and relics in the GMRT
cluster sample are located in clusters with some sign of disturbance. 
Conversely, looking at the results from the point of view of the cluster 
dynamical status, it is noticeable that none of the relaxed clusters hosts 
halos or relics, while unrelaxed clusters may or may not host a diffuse 
source. 
More recently, Cassano et al. (2010a) 
cross--checked the presence of a radio halo (and lack thereof) for a subsample 
of the GMRT radio halo cluster sample with the cluster dynamical status,
derived by means of three different quantitative estimators on the basis 
of high quality X--ray {\it Chandra} imaging. 
The results clearly show that the cluster 
radio bimodality has a correspondence in terms of cluster dynamics: radio 
halos are found in dynamically disturbed systems, while "radio quiet" clusters 
are more relaxed. Their study includes also a few mini--halo clusters 
(see Section 2.2), which are confirmed to be found in relaxed systems, 
as already known for the historical mini--halos, such as Perseus 
(A\,426) and A\,2052 
(Fabian et al. 2006 and Blanton et al. 2001 respectively). 

\section{Spectra of radio halos}

Spectra of radio halos contain important information related to their
origin.
Imaging of the spectral index distribution provides information on the
energy spectrum of the radiating electrons and on the magnetic field 
distribution. However, only for a handful of halos such observational 
information is available, mainly due to the decreasing sensitivity of low 
frequency imaging. A patchy distribution seems to be a common feature,
as is observed in for instance in A\,2744 \citep{orru07} and in
A\,3562 \citep{giacintucci05}. In other cases, such as Coma
\citep{giovannini93}, a steepening is observed at increasing distance 
from the cluster centre.

The accurate measurement of the integrated spectra of radio halos is a 
difficult task. Radio halos usually embed a number of individual sources, 
whose flux density needs to be carefully subtracted from the total diffuse 
emission, and this requires high quality imaging over a range of resolutions.
Moreover, diffuse cluster sources are best imaged at low frequency, and 
high quality imaging at frequencies below 1.4 GHz has become available
only very recently. For this reason, a spectrum 
with at least three datapoints spread over $\sim$ 1 order of 
magnitude is available only for few objects . 
Beyond the well--known case of Coma--C (Thierbach et al. 2003), 
good spectra spectra are
available for A\,2256 \citep{brentjens08}, 
RXCJ\,2003.5-2323 \citep{giacintucci09a}, 
A\,521 \citep{brunetti08,dallacasa09}, 
A\,697 \citep{macario10}, 
A\,3562 \citep{venturi03}. 

%
\begin{figure}[htbp!]
\resizebox{\hsize}{!}{\includegraphics[clip=true]{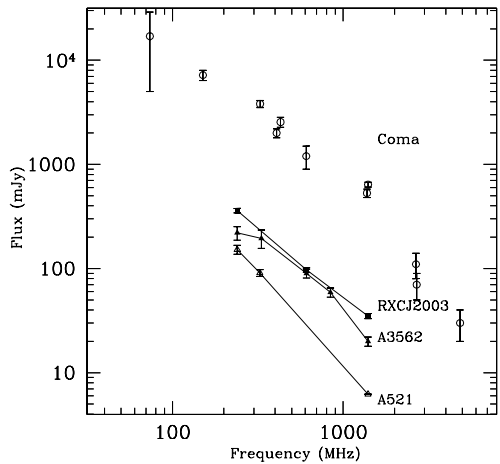}}
\caption{\footnotesize
\textit Spectra for a few radio halos are reported. Empty triangles: A\,521 
(Dallacasa et al. 2009); filled triangles: A\,3562 (Giacintucci et al. 2005)
filled squares: RXCJ\,2003.5--2323 (Giacintucci et al. 2009a); 
empty circles: Coma (Thierbach et al. 2003).}
\label{fig:fig4}
\end{figure}

It is now becoming clear that there is not a ``typical spectrum'' for radio 
halos and that there is a considerable spread in the spectral index values. 
Moreover, while spectra based on few measurements tend to
introduce a bias towards a single power--law shape, those cases with
better frequency sampling show a high frequency cutoff
(for $\nu_b \ge$ 1 GHz, i.e. Coma, Thierbach et al. 2003) and a low frequency 
flattening (for $\nu \le$ few hundred MHz, i.e. A\,3562 and A\,521, Venturi et 
al. 2003 and Brunetti et al. 2008 respectively).
Fig. \ref{fig:fig4} shows four meaningful examples. Apart from the flux density
scale, it is clear that the spectra shown are all very different.

\subsection{Ultra--steep spectrum radio halos}
Beyond the historical sources, most of the radio halos known so far were 
found either by inspection of relatively shallow surveys, such as the 1.4 GHz 
Northern VLA Sky Survey (NVSS) and the 327 MHz Westerbork Northern Sky Survey 
(WENSS) (\citealt{giovannini99} and\citealt{ks01} respectively), leading
to the detection of relatively bright sources at GHz frequencies, which
do have spectral index values in the range 1.2--1.4, as often reported in the
literature.
The 610 MHz GMRT Radio Halo Survey, roughly 5 times more sensitive than the
NVSS, allowed the detection of considerably fainter sources, at the very
limit of detectability on the NVSS, and led to the discovery of a new 
``population'' of radio halos, with very steep spectrum at GHz frequencies, 
i.e. $\alpha\sim 2$.
A\,521 is the first radio halo found with spectral index $\alpha\sim 2$ 
\citep{brunetti08,dallacasa09}:
the images published in the literature show that the source becomes
bright and easily detectable at frequencies of the order of 325 MHz and below,
and that the relative dominance of the halo and of the relic change 
considerably going to lower frequencies, as a result of the different
spectral shapes of the two sources (for the relic 
$\alpha=1.48$,\citealt{giacintucci08}). Another very steep spectrum 
halo was found in A\,697 (\citealt{macario10}, Macario et al. this volume),
as well as few candidates currently under investigation.
Both A\,521 and A\,697 show that the size of the radio halo increases
with decreasing frequency, which is different from what is observed in
the ``classical'' radio halos. For instance, the shape and extent of
RXCJ\,2003.5--2323 (Fig. \ref{fig:fig3}) does not change at least in the
range 235 MHz -- 1.4 GHz. The same is true also for Coma (but 
see~\citealt{brown11}).

The finding of very steep spectrum halos is opening up a new window in
our understanding of the origin and formation of radio halos and on the 
radio halo--cluster merger connection. One of the consequences of the 
turbulent re--acceleration model (see Brunetti and Cassano, present volume)
is the dependance of the cutoff frequency $\nu_b$ on the efficiency of
the re--acceleration process during the cluster mergers: less efficient 
re--acceleration shifts $\nu_b$ to lower and lower frequencies.
An application of the turbulent re--acceleration model \citep{cassano10b}
shows that a population of ultra--steep spectrum radio halos is expected, 
as due to to less energetic mergers and/or less massive clusters.

These results suggest that the spectra of radio halos may be a key information 
for our understanding of the dynamical processes at play in the hosting 
clusters.

\section{Relics and shocks}

Radio relics show a broad variety of morphologies, and a considerable 
spread in size and projected distance from the cluster centre, going 
from A\,3667 and Coma, whose relics are beyond the 1.5 Mpc size and are 
located more than 1.5 Mpc from the cluster centre (Rottgering et al. 1997 and
Brown \& Rudnick 2011 respectively) to the small relic in A\,4038 
\citep{kale10}. An updated
summary of our knowledge of the observational properties of relics, as
well as significant statistical correlations, is given in
van Weeren et al. (2009a).

The location, shape and polarization properties of relics suggest that their 
origin is connected to the propagation of shock waves, with Mach numbers 
\ltsim 3, during cluster mergers. Indeed all models proposed so far for 
the formation of relics require the presence of a shock at the relic position,
which may (re--)accelerate relativistic electrons or ``revive'' fossil
radio plasma by means of adiabatic compression 
\citep{ensslin98,ensslin01,markevitch05}.
For a few well studied cases, the spectrum of the radio halo is consistent
with Mach numbers for the shock in the range 1--3 (i.e. 
A\,521,~\citealt{giacintucci08}; CIZA\,J2242.8+5301,\citealt{vw10}; 
A\,754,~\citealt{macario11}).

So far only a few shocks have been firmly detected with {\it Chandra}
(Markevitch et al., present volume), and it would
be crucial to increase the statistics not only of confirmed shocks, but also
of clusters with a clear connection between a shock front and a relic, 
as it has been
the case for A\,520 \citep{markevitch05} and A\,754 \citep{macario11}.

\section{Open questions and future perspectives}

Our knowledge and understanding of diffuse cluster sources, and their 
connection to the processes of cluster assembly in the Universe has
considerably improved over the past few years. It seems now firmly
established that cluster mergers are at the origin of radio halos and relics,
but many details need to be clarified and understood. 
A few points deserve special attention, and these are 
summarized below.
\\
{\it (1)} High sensitivity imaging of radio halos at low frequencies is
crucial. So far only two radio halos are confirmed ultra--steep spectrum
sources. It is essential that more objects are found, to safely
talk of a ``new population''. Finding more USSRH would
provide further observational support in favour of the turbulent 
re--acceleration model. According to the statistical results of 
Cassano et al. (2010b), such objects should open up the study of 
the observational effects of the common and numerous minor mergers,
expected to be the dominant mechanisms of mass assembly in the Universe.
\\
{\it (2)} It is essential to increase the number of radio halos and
relics with well defined integrated spectra, since the spectral shape
and spectral index value are the signature of the underlying electron
population and of the re--acceleration processes.
\\
{\it (3)} Imaging of the spectral index distribution is another crucial piece 
of information which we still miss. It is relevant not only to address 
point (2), but also for the study the distribution of the magnetic field in 
the cluster volume.
\\
{\it(4)} The number of clusters with multiple diffuse sources is
steadily increasing, especially after the advent of the good
quality images provided by the GMRT at low frequencies (see for instance
van Weeren at al., present volume). This is most likely telling us
something about the development of shocks and turbulence in the
cluster volume as consequence of cluster mergers, and their signature
on the diffuse radio emission.

The future is bright. LOFAR has started to deliver the first
images (see Rottgering, present volume), and it is expected to unveil
the population of ultra--steep spectrum halos which the GMRT has just
discovered. The few $\mu$Jy sensitivity of the EVLA will lead to the detection
of very faint halos at GHz frequencies, thus constraining the bright end of 
the radio spectrum of ultra--steep sources. At the same time, the GMRT still
has a long way ahead to continue its observational contribution in this 
field.
To conclude, I believe that a new era in the study of diffuse cluster
sources is about to start.


\begin{acknowledgements}
I acknowledge my collaborators, S. Giacintucci, G. Brunetti, R. Cassano,
D. Dallacasa, G. Macario and R. Athreya. 
I thank GMRT staff, in particular N. Kantharia, for their help and assistance
throughout the observations of the GMRT Radio Halo Survey and its 
follow--up.
The GMRT is run by the National Centre for Radio Astrophysics of the Tata 
Institute of Fundamental Research. 
This work is partially supported by PRIN INAF 2008 and ASI I/088/06/0. 
\end{acknowledgements}

\bibliographystyle{aa}

\end{document}